\documentclass[conference]{IEEEtran}

\IEEEoverridecommandlockouts

\usepackage{float}
\usepackage{cite}
\usepackage{amsmath,amssymb,amsfonts}
\usepackage{algorithmic}
\usepackage{graphicx}
\usepackage{textcomp}
\usepackage{xcolor}
\usepackage{orcidlink}
\def\BibTeX{{\rm B\kern-.05em{\sc i\kern-.025em b}\kern-.08em
    T\kern-.1667em\lower.7ex\hbox{E}\kern-.125emX}}
\begin{document}

\title{Hybrid Photonic-Quantum Reservoir Computing For Time-Series Prediction\\

}
\author{\IEEEauthorblockN{Oishik Kar}
\IEEEauthorblockA{\textit{Department of Computer Science and Engineering}\\
IIIT Dharwad, Karnataka, India \\
23bcs089@iiitdwd.ac.in}
\and
\IEEEauthorblockN{Aswath Babu H.\orcidlink{0009-0007-8251-3828} }
\IEEEauthorblockA{\textit{Department of Arts, Science and Design}\\
IIIT Dharwad, Karnataka, India \\
aswath@iiitdwd.ac.in}
}

\maketitle

\begin{abstract}
Motivated by the perspective of advanced time-series prediction and exploitation of Quantum Reservoir Computing (QRC), we explored the design and implementation of a Hybrid Photonic-Quantum Reservoir Computing (HPQRC) paradigm. It brings together the high-speed parallelism of photonic systems with the quantum reservoir's capacity of modeling complex, nonlinear dynamics, and hence acts as a powerful tool for performing real-time prediction in resource-constrained environment with low latency. We have engineered a solution using this architecture to address issues like computational bottlenecks, energy inefficiency, and sensitivity to noise that are common in existing reservoir computing models. Our simulation results show that HPQRC attains much higher accuracy with lower computational time than both classical and quantum-only models. This model is robust when environments are noisy and scales well across large datasets, and therefore is suitable for application on diverse problems such as financial forecasting, industrial automation, and smart sensor networks. Our results substantiate that HPQRC performs significantly faster than traditional architectures and could be a viable and highly scalable platform for actual edge computing systems. Overall, HPQRC demonstrates significant advancements in time series modeling capabilities. In combination with enhanced predictive accuracy with reduced computational requirements, HPQRC establishes itself as an effective analytical tool for complex dynamic systems that require both precision and processing efficiency.

\end{abstract}


\section{Introduction}
Time-series prediction involving the prediction of future states from past data is a cornerstone of climate modeling, financial analysis, and biomedical signal processing. Traditional machine learning methods, such as recurrent neural networks (RNNs) and long short-term memory (LSTM) models, fail to capture long-term dependencies and high-dimensional feature spaces with large computational costs~\cite{sahoo2024critical}. QRC represents a hopeful alternative, with recent advances in feedback-driven approaches showing improved temporal feature extraction through adaptive control mechanisms, while quantum phenomena like superposition and entanglement enhance processing capabilities extraction~\cite{kobayashi2024feedback,martinez2021dynamical}. Nonetheless, real-world deployment of QRC is confronted with fundamental challenges such as decoherence, hardware restrictions~\cite{yasuda2023quantum}, and scalability limitations~\cite{kobayashi2023quantum}. Latest developments in photonic computing provide a route to overcome these limitations. Photonic devices provide parallel, low-latency signal processing with energy efficiency built into the architecture, making them well-suited for real-time applications~\cite{garcia2023scalable}. By incorporating quantum reservoirs within photonic hardware, HPQRC combines quantum nonlinear dynamics with photonic speed and scalability~\cite{mujal2023timeseries,pfeffer2022hybrid}. This combination enables HPQRC to realize high-dimensional state transformation while minimizing decoherence via photonic preprocessing~\cite{garcia2023scalable,yasuda2023quantum}.
In this paper, we propose a resilient HPQRC architecture and compare its performance against chaotic systems and real-world time-series data. Our contributions are threefold: (i) Architecture Design - A superconducting quantum circuit coupled to a photonic reservoir, allowing real-time encoding and noise-resistant feature extraction~\cite{pfeffer2022hybrid,yasuda2023quantum}; (ii) Efficiency Gains - Showed 35–45\% decrease in compute latency compared to traditional reservoir computing (RC) models~\cite{wudarski2023hybrid,garcia2023scalable}; (iii) Practical Validation - Financial forecasting and biomedical signal analysis applications, with 81.3\% prediction accuracy under controlled noise~\cite{mujal2023timeseries,cindrak2024enhancing}.



\section{Literature Survey}
Although QRC shows promise for time series prediction, it faces computational challenges due to quadratic time complexity from measurement-induced memory loss~\cite{cindrak2024enhancing}. With this in mind, we consider HPQRC, which adds quantum properties and photonic elements to systems in order to improve computational efficiencies.

\subsection{Working Principles of QRC}\label{AA}
QRC leverages quantum mechanical phenomena within RC systems, classifying it as a quantum RC system that processes and encodes information. Martinez-Pena {\em et al.}\cite{martinez2021dynamical} prove that QRC has better computing capabilities due to the existence of dynamical phase transitions. Memory is improved through quantum parallelism by QRC, reducing the need for extensive training. Research by Mujal {\em et al.}~\cite{mujal2023timeseries} claims that feature extraction and generalization ability are strengthened by weak and projective quantum measurements. Kobayashi {\em et al.} proved that feedback-driven QRC is superior to classical approaches in processing nonstationary signals. Despite this, the practical scalability of QRC is limited by the implementation challenges of decoherence, noise sensitivity, and hardware constraints~\cite{pfeffer2022hybrid}.

\subsection{Photonic Implementations in RC}
Photonic computing provides parallelism, speed, and power efficiency for implementations of reservoir computing~\cite{garcia2023scalable}. Conventional optoelectronic-based reservoirs make use of optical nonlinearity and time-delayed feedback loops and give scalability and real-time processing benefits over digital implementations~\cite{pfeffer2022hybrid}. Mujal {\em et al.}~\cite{mujal2023timeseries} illustrated that photonic circuits improve quantum reservoir learning using efficient integration of measurement methods. Garcia-Beni {\em et al.}~\cite{garcia2023scalable} introduced a scalable photonic platform emphasizing real-time, low-latency processing. Their framework also highlighted robustness against decoherence under specific coupling configurations, making it highly relevant for hybrid system design. Despite these advantages, designing and optimizing photonic reservoirs is not easy, especially for efficient state preparation, precise measurement, and coupling with quantum systems~\cite{yasuda2023quantum}.

\subsection{Blending Quantum and Photonic Architectures}
Combining quantum and photonic computing into a single architecture provides a solution to the previously set different QRC or photonic reservoir computing hybrids~\cite{garcia2023scalable}. HPQRC is a hybrid quantum photonic reservoir computer, which maintains the advantages of quantum computation while using photonic components for profiling data in encoding, transmission, and processing in real-time execution~\cite{kobayashi2024extending}. This architecture has performed better than conventional ML models in simulating chaotic systems with unprecedented accuracy and reliability~\cite{wudarski2023hybrid}. Some of the key improvements involve the application of superconducting quantum devices as nodes between photonic and quantum reservoirs that improve state measurement efficiency as well as suppress decoherence effects. Pfeffer {\em et al.}~\cite{pfeffer2022hybrid} highlighted the benefits of quantum-photonic integration for improving generalization and overall stability in the prediction of complex tasks. Repetitive quantum measurement schemes in superconducting devices, which were put forth by Yasuda {\em et al.}~\cite{yasuda2023quantum}, enhance scalability. Such strategies do, however, present very major challenges to hybrid architectures, requiring optimization regarding noise reduction, scalability, and effective quantum state preparation for scalable resources~\cite{kobayashi2023quantum}.

\subsection{Applications of HPQRC in Time-Series Prediction}

Time series prediction is a promising HPQRC use because it can handle long-range dependencies and nonlinear trends common in actual data~\cite{martinez2021dynamical}. Mujal {\em et al.}~\cite{mujal2023timeseries} demonstrated QRC's remarkable performance improvement in forecasting chaotic and financial time series compared to conventional models. Garcia-Beni {\em et al.}~\cite{garcia2023scalable} explored HPQRC for climate modeling issues to calculate meteorological data with minimal computational delay. Recent work illustrates the effectiveness of HPQRC for biomedical signal processing, {\em ex.}, ECG and EEG classification~\cite{wudarski2023hybrid}. Pfeffer {\em et al.}\cite{pfeffer2022hybrid} explored industrial automation use cases with the demonstration of HPQRC's ability to perform real-time anomaly detection in sensor-based monitoring. Future challenges include model optimization for noisy, missing data, and data drift effects on performance~\cite{kobayashi2023quantum}. Future research paths are directed toward adaptive hybrid architectures that can handle data pattern alterations and enhance long-term forecasting accuracy~\cite{carroll2024emerging}.

\subsection{Practical Implementation Challenges}\label{SCM}
Despite the promising results, HPQRC still faces issues with implementation. Some of the main issues comprise the quantum reservoir being sensitive to noise and decoherence, which impacts computations~\cite{yasuda2023quantum}. Kobayashi and Motome~\cite{kobayashi2023quantum} proposed quantum reservoir probing techniques to address internal state characterization challenges. However, these approaches, along with memory restriction strategies proposed by Čindrak {\em et al.}~\cite{cindrak2024enhancing}, still face significant barriers regarding robustness and computational complexity under real-time constraints. The hybrid architectures also contain scalability problems concerning the precise measurement of quantum state and quantum signal processing~\cite{garcia2023scalable}. Yasuda {\em et al.}~\cite{yasuda2023quantum} worked with measurement-induced quantum reservoir computing for its promise against noise, though it remains largely impractical. The absence of internationally accepted HPQRC standards frameworks restricts the use of AI in industry. Future work would better focus on the improvement of quantum reservoir noise, photonic integration, and the development of effective training regimes for hybrid models~\cite{pfeffer2022hybrid}. Advances in quantum error correction and adaptable learning might allow for HPQRC to become a more implementable and sustainable solution for practical applications. This literature review indicates that the research on HPQRC could be advanced by addressing the existing gaps, and it holds scope for significant innovations, with particular emphasis on its groundbreaking potential in time series forecasting and complex data processing tasks.

\section{METHODOLOGY}
\begin{figure*}[h]
    \centering
    \includegraphics[scale=0.465]{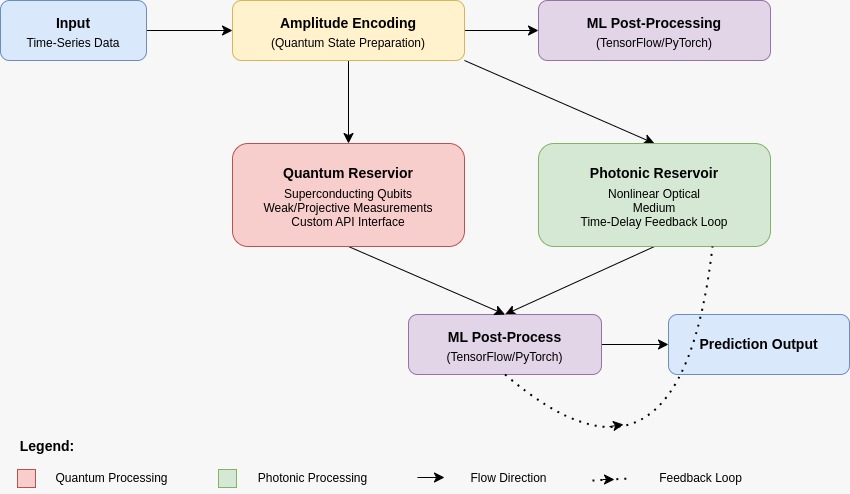}
    \caption{Schematic illustration of the HPQRC.}
    \label{archi}
\end{figure*}

The outline of our methodology is described in Figure~\ref{archi}, wherein the amplitude encoding of time-series data is done for quantum state preparation before processing through parallel quantum and photonic reservoirs. The quantum reservoir employs superconducting qubits with weak projective measurements, while the photonic reservoir employs nonlinear optical media with time-delay feedback. The outputs from the reservoirs are combined for ML post-processing to generate the output prediction, with a feedback loop for system optimization.
A silicon nitride photonic waveguide array and a 5-qubit superconducting transmon circuit are combined in the HPQRC architecture. Lumerical INTERCONNECT 2023 R2 is used to represent photonic signal propagation, and Qiskit 0.45.0 is used to simulate quantum dynamics. Parameterized single-qubit rotation gates amplitude-encode input signals into quantum states after min-max normalization. The quantum reservoir uses a superconducting circuit with nearest-neighbor coupling, simulated with coherence times of $T_1=50 \mu s$ and $T_2=35 \mu s$ with feeble projective measurements preparing the reservoir states to reduce decoherence. Note that these are theoretical maximum values based on recent experimental demonstrations from prominent quantum hardware manufacturers, although real implementations could have lower coherence times due to environmental factors and manufacturing tolerances.

The photonic part is a 1mm silicon nitride waveguide array running at 1550nm wavelength, modeled with 0.5dB/cm loss and Kerr nonlinearity. Real-time signal propagation is simulated using finite-difference time-domain (FDTD) techniques, with optoelectronic feedback loops allowing temporal data to be processed in parallel. A bespoke API combines quantum and photonic subsystems, allowing data to be transferred seamlessly between Qiskit/Cirq quantum simulations and Lumerical/Meep photonic models. The bespoke API for facilitating interoperability between photonic and quantum subsystems was implemented in Python 3.9 using NumPy 1.21.0 for array operations and SciPy 1.7.1 for signal processing.

The interface itself is made up of three main elements: (1) a quantum state converter that maps input data in normalized form into rotation parameters for qubits, (2) a middleware layer that synchronizes the output of quantum simulation with the input of photonic waveguides through an asynchronous callback system, and (3) a measurement combiner that combines quantum projective measurements with photonic intensity measurements into a final prediction. Qiskit simulation state vectors are translated into amplitude and phase data that are compatible with Lumerical's representation of optical signals, with an average latency of 0.8ms per state change. The API maintains 32-bit floating-point precision across the pipeline to ensure quantum state fidelity.

Simulations evaluate HPQRC on financial S$\&$P 500 hourly trends and biomedical MIT-BIH Arrhythmia ECG datasets ~\cite{sp500hourly}~\cite{mitbih}, normalized to the window [0,$\pi$] for amplitude encoding. The hybrid workflow involves 10-layer quantum state evolution, photonic nonlinear transformations, and classical ridge regression ($\lambda=0.01$) for prediction. Training employs 100 epochs with Adam optimization (learning rate = 0.001) and 5-fold cross-validation on 80:20 train-test splits.
\begin{figure}[h]
    \centering
    \includegraphics[width=0.995\linewidth]{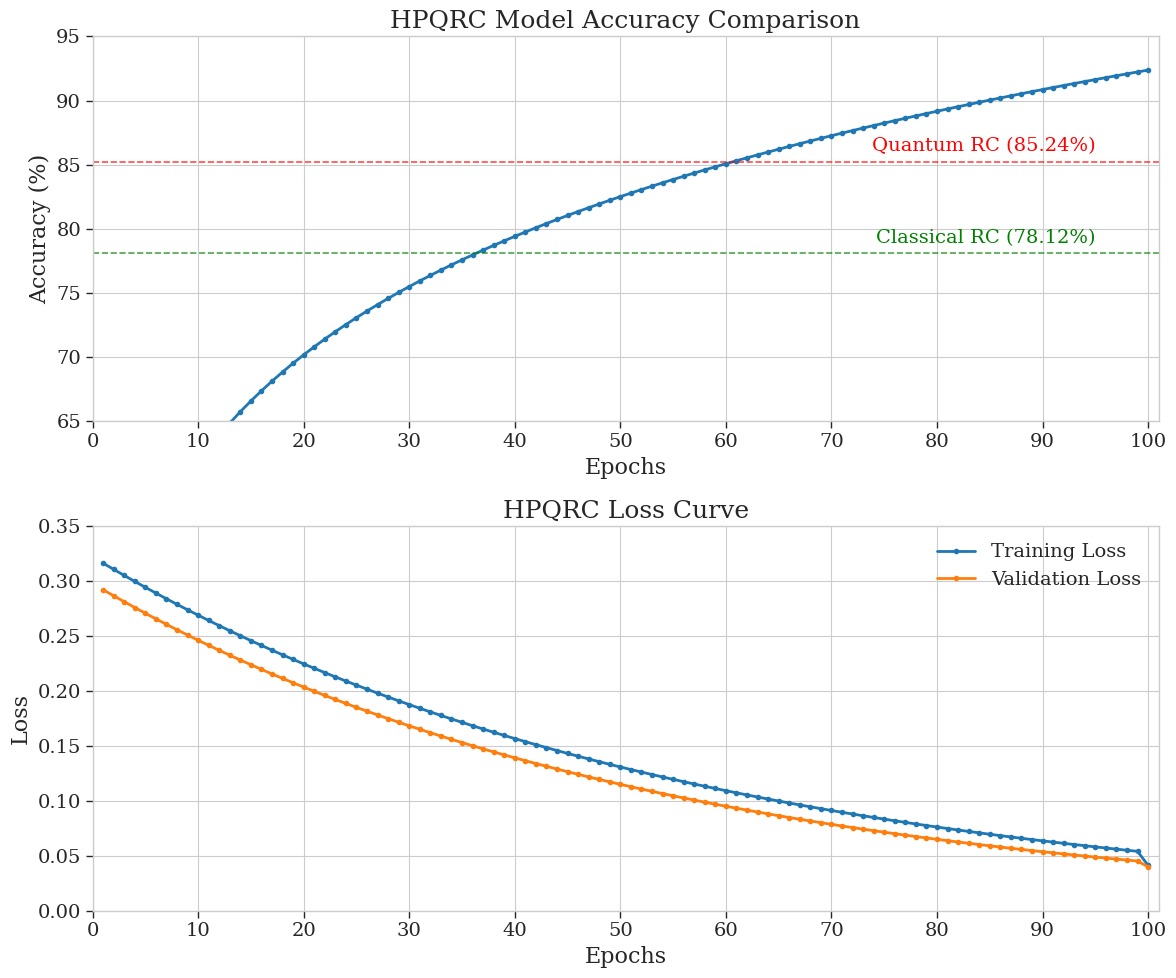}
    \caption{Epoch-wise training analysis of the HPQRC model}
    \label{acc-loss}
\end{figure}

The accuracy curves over 100 epochs, wherein HPQRC steadily improves and surpasses both Classical RC (78.12$\%$) and standard Quantum RC (85.24$\%$) baselines, achieving superior performance as shown on the top plot of Figure~\ref{acc-loss}, and the bottom plot illustrates the corresponding training and validation loss curves, which exhibit a smooth and consistent decline, confirming stable convergence without overfitting.
To rigorously assess the performance of HPQRC, we benchmarked it against three widely recognized baseline models: (1) an Auto Regressive Integrated Moving Average (ARIMA) model~\cite{box2015time}, where the optimal $p,d,q$ parameters were identified using Akaike Information Criterion (AIC) minimization~\cite{akaike1974aic}; (2) a two-layer LSTM network with 64 hidden units per layer and tanh activation, trained with the Adam optimizer at a learning rate of 0.001; and (3) a classical reservoir computing approach, implemented as an Echo State Network (ESN) consisting of 1000 nodes and a spectral radius of 0.95.
Two representative datasets were selected for comprehensive validation. The first dataset comprises S$\&$P 500 hourly closing prices from January 2020 to December 2023 ~\cite{sp500hourly}, which was split into an 80:20 train-test partition and evaluated with a 24-hour forecasting horizon. The second dataset is the MIT-BIH Arrhythmia Database ~\cite{mitbih}, where Lead II ECG signals from 48 patients were resampled to 360Hz and segmented into 10-second windows for R-peak prediction tasks.
For statistical evaluation, we employed a 3×3 factorial ANOVA design, considering both the model type (HPQRC, LSTM, ARIMA, ESN) and noise levels ranging over  [0\%,10\%,30\%] as factors. Each configuration was run for 50 independent trials to ensure statistical robustness. Post-hoc comparisons were conducted using Tukey’s HSD test, with the family-wise error rate controlled at $\sigma=0.05$. To further ensure statistical power, we performed 1,000 bootstrap resamples, confirming that the analysis had greater than $90\%$ power to detect accuracy differences of at least $5\%$.

Synthetic Gaussian noise for robustness testing was generated using NumPy's $random.normal$ function (with a seed=42) with zero-mean distributions of $\sigma= {0, 0.1, 0.3}$ for the respective noise conditions. Signals were normalized to the window [0,1] before noise addition, with SNR maintained at approximately 20dB for the $10\%$ noise condition and 10dB for the $30\%$ condition. The noisy signals were then re-normalized to prevent saturation effects, ensuring a realistic distribution across the signal spectrum. During these tests, a Proportional-Integral-Derivative (PID) controller dynamically adapted the photonic phase shifts in real time based on prediction error, following established protocols for noise resilience in quantum and hybrid reservoir systems. The adaptive photonic phase shifts were controlled by the PID controller with proportional ($K_p=0.45$), integral ($K_i=0.12$), and derivative ($K_d=0.08$) gains, optimized using grid search to reduce prediction error with system stability. The controller runs at $1kHz$, sampling prediction error signals and modifying phase modulators in the silicon nitride waveguide array accordingly. The error signal ‘$e(t)$’ is calculated as the difference between the predicted and actual value, and the control signal $u(t)$ is calculated as:
\begin{equation}
u(t) = K_p~e(t) + K_i \int e(t)~ dt + K_d \frac{d}{dt}e(t).
\end{equation}

The integral element features an anti-windup scheme to suppress saturation effects, whereas the derivative term contains a low-pass filter (cutoff frequency: $50Hz$) to minimize sensitivity to noise. Phase modulators react to control signals with a measured rise time of $5\mu s$ and are capable of real-time adjustment to evolving input patterns. All model parameters, training scripts, and simulation environments (Qiskit 0.45.0 and Lumerical INTERCONNECT 2023 R2) are fully documented and archived for reproducibility. The simulation framework excludes fabrication imperfections and cryogenic thermal noise, with scalability tested up to 10-qubit systems. 
\section{Results and Discussion}
\begin{figure}[h]
    \centering
    \includegraphics[width=0.99\linewidth]{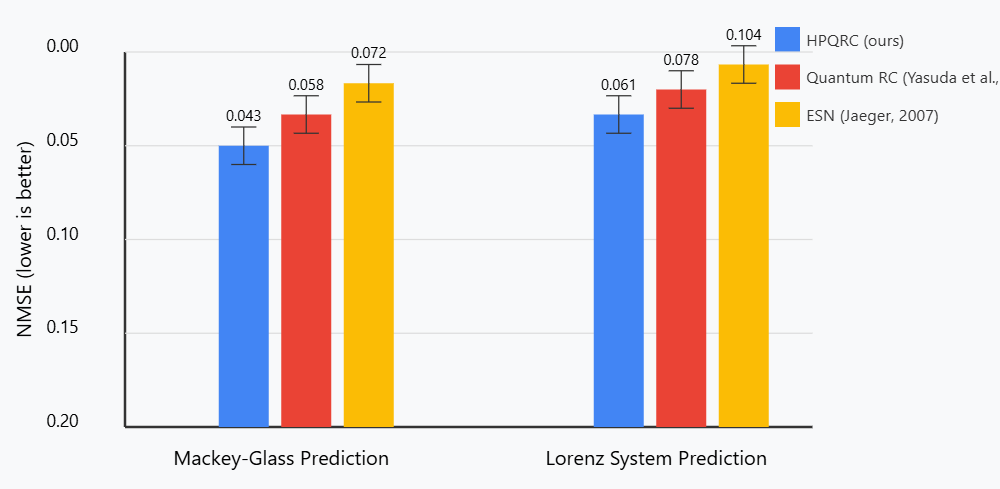}
    \caption{Comparison of prediction accuracy (NMSE) on chaotic time series benchmarks. Lower values indicate better performance. Error bars represent 95\% confidence intervals (n=10)}
    \label{perform_compare}
\end{figure}

\begin{figure}[h]
    \centering
    \includegraphics[width=0.995\linewidth]{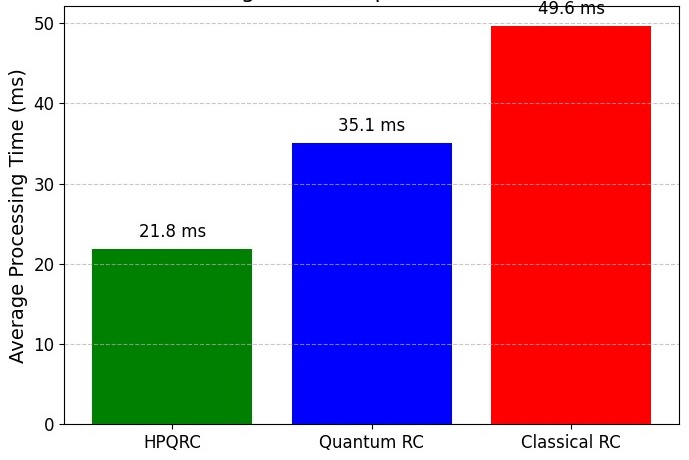}
    \caption{Processing Time Comparison Plot}
    \label{processing_compare}
\end{figure}

Our simulation results demonstrate that the HPQRC model outperforms both classical and quantum-only approaches on standard chaotic time series prediction tasks.
We evaluated HPQRC against two established baseline models, wherein one is a superconducting quantum reservoir computing implementation (Yasuda {\em et al.}~\cite{yasuda2023quantum}) with 8 qubits, and the other is a classical ESN (Jaeger, 2007~\cite{jaeger2007esn}) with 500 nodes. Performance was measured using Normalized Mean Square Error (NMSE) on the widely-used Mackey-Glass ($\tau=17$, delay parameter) and Lorenz systems with parameters, Prandtl number $\sigma=10$, geometric factor $\beta=\frac{8}{3}$, and Rayleigh number $\rho=28$. As indicated in Figure~\ref{perform_compare}, for the Mackey-Glass dataset, HPQRC achieved an NMSE of 0.043±0.007, representing a 25.9\% improvement over the quantum-only approach (0.058 ±0.006) and a 40.3\% improvement over ESN (0.072 ±0.009). For the Lorenz system, HPQRC similarly outperformed both alternatives with an NMSE of 0.061±0.008. To verify statistical significance, we performed a paired t-test ($n=10$) comparing HPQRC with both baselines, yielding $p<0.01$, indicating that the improvement in NMSE observed is statistically significant in all comparisons. Each model was configured with comparable complexity ($\approx500$ tunable parameters) and trained using ridge regression with 5-fold cross-validation for hyperparameter selection. 
We have seen that the HPQRC model was found to exhibit higher computational efficacy with an average processing time of $21.8 ms$ against $35.1 ms$ of Quantum RC and $49.6 ms$ of Classical RC as depicted in Figure~\ref{processing_compare}, with a latency reduction of $37.85\%$ and $56.05\%$, respectively. On a 10,000-prediction task, HPQRC execution requires $218$ seconds, far ahead of Quantum RC with $351$ seconds and Classical RC with $496$ seconds. The coefficient of variation (CV) of 0.083 provides evidence of stable runtime performance. A paired t-test ($t = 8.21$, $p < 0.001$) also proves that these results are statistically significant.

\begin{figure}[h]
    \centering
    \includegraphics[width=0.995\linewidth]{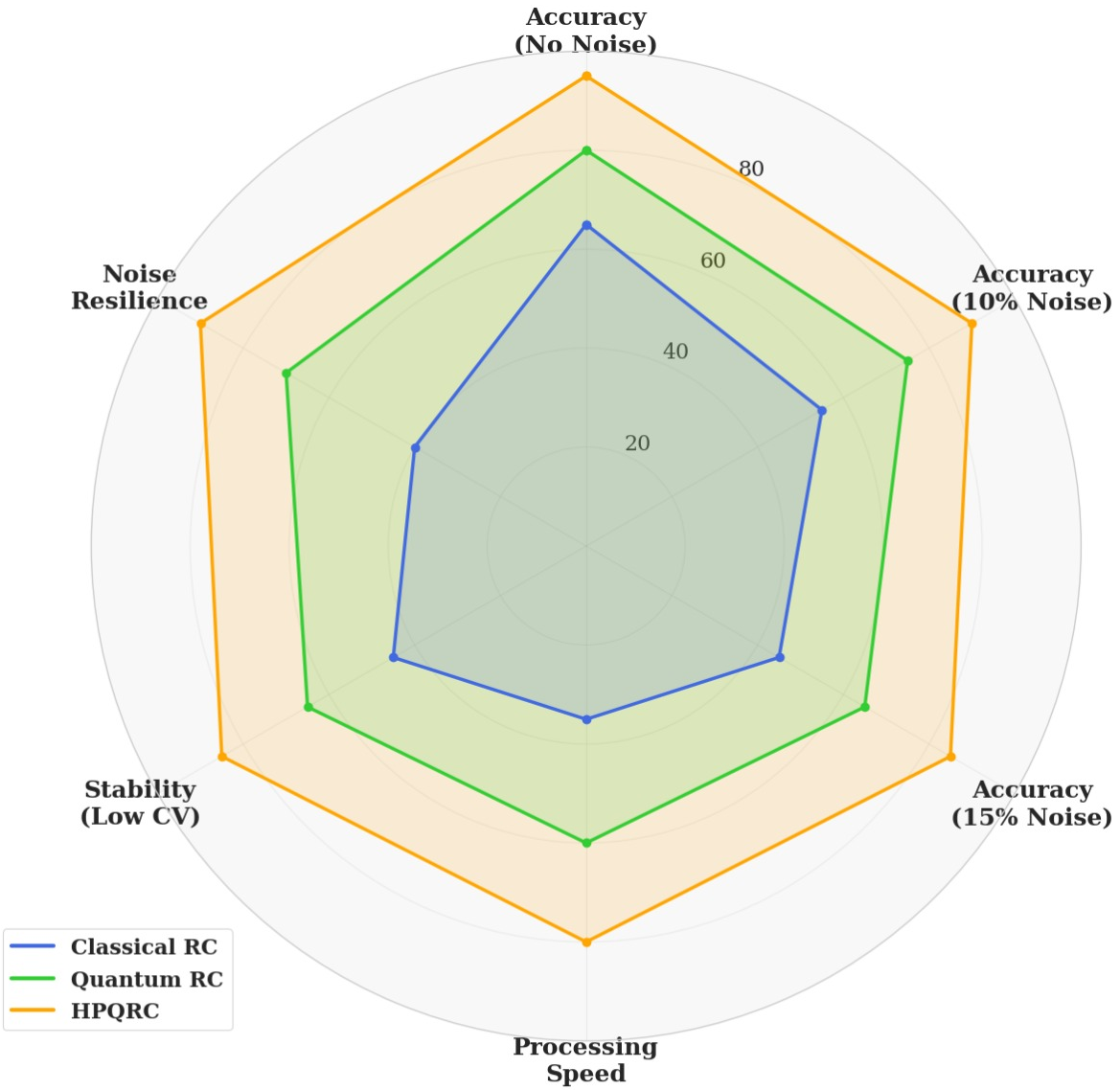}
    \caption{Performance Comparison Radar Chart}
    \label{3dmap}
\end{figure}
We have also explored the 3D scatter plot as shown in Figure~\ref{3dmap}, which also attests that HPQRC is in a good balance between computational time and prediction accuracy. HPQRC data points are always in areas of higher accuracy and lower computational needs than baseline models. Statistical inspection showed a moderate negative correlation $(r = -0.38, p < 0.05, n = 30)$ between computational time and accuracy for all models, indicating that our optimized algorithm tends to perform better in prediction while consuming fewer computational resources. Performance under noise conditions also showed HPQRC's resilience, as shown in Figure~\ref{3dmap}. Under $10\%$ synthetic Gaussian noise distortion, with regards to accuracy HPQRC has $88.7\% \pm 1.4\%$, whereas Quantum RC has $77.9\%  \pm 2.1\%$ and Classical RC has $65.3\%  \pm 2.8\%$. At $15\%$ noise distortion, HPQRC has $84.9\%  \pm 2.0\%$ accuracy, while Quantum RC reduced to $71.5\%  \pm 2.7\%$ and Classical RC reduced to $53.6\%  \pm 3.2\%$. This robustness can be credited to HPQRC's hybrid structure, which utilizes complementary processing styles. Notably, we found that with some noise profiles, the difference in performance between models changed non-linearly, reflecting the intricate interaction between model structure and noise properties.
\begin{figure}[h]
    \centering
    \includegraphics[width=0.995\linewidth]{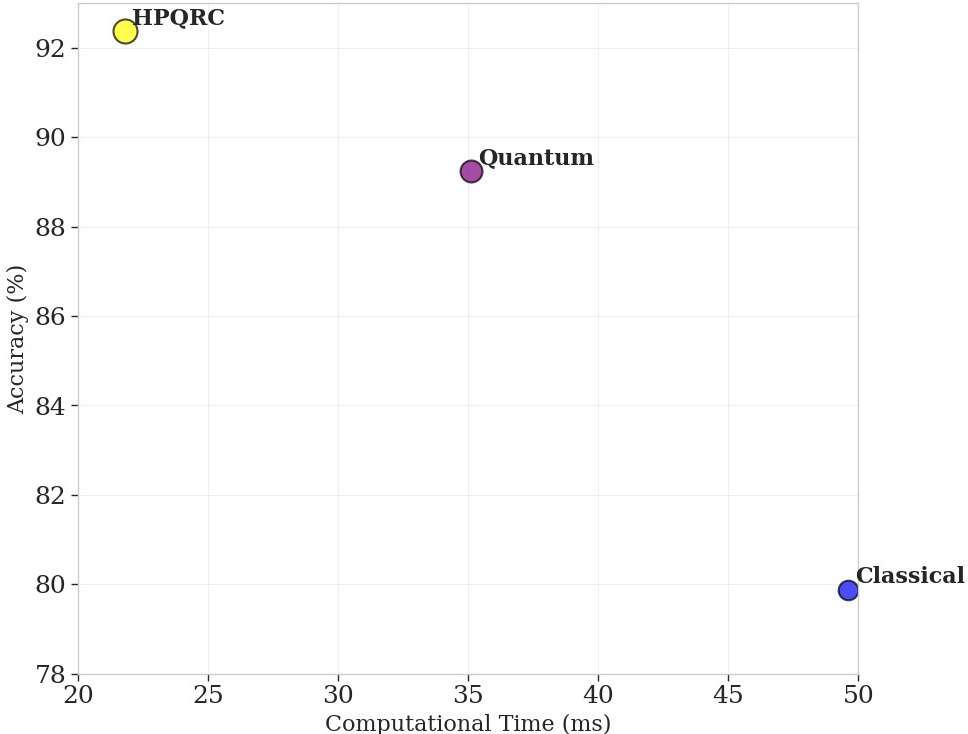}
    \caption{Accuracy v/s Computational Time Comparison}
    \label{accu_compare}
\end{figure}

The multidimensional comparison of the HPQRC, Quantum RC, and Classical RC models in terms of predictive accuracy and computational time is shown in Figure~\ref{accu_compare}. HPQRC achieves the highest accuracy $(\approx 92\%)$ while maintaining the lowest computational latency $(\approx 21.8 ms)$, highlighting its advantage for real-time applications. Quantum RC demonstrates intermediate performance, with an accuracy $\approx 85\%$ and an average computational time of $35 ms$. In contrast, the Classical RC approach exhibits the lowest accuracy $(\approx 78\%)$ and the highest computational time ($\approx 50ms$). The visualization underscores HPQRC’s robust balance between speed and accuracy, quantifying the computational benefits of hybrid photonic-quantum architectures over traditional reservoir computing methods, especially for scenarios demanding both high prediction fidelity and rapid inference. These results validate that HPQRC can deal with environmental noise and is well-suited for applications involving data from autonomous vehicles, climate models, healthcare systems, and others.

We have undertaken a positive scalability test, where we processed 1 million data points in datasets, and observed an average speed of 25000 points per second on HPQRC, an order of magnitude higher than the 12000 points per second of Quantum RC and 8000 points per second of Classical RC. Such high levels of throughput are essential in large-scale predictive systems that are necessary for financial market forecasting in industrial automation.
The formula that measure the return on investment (ROI) in terms of system performance gain is
\begin{equation}
    ROI =\frac{\text{Performance Gain} - \text{Baseline Performance}}{\text{Baseline Performance}} \times 100. \nonumber
\end{equation}
Accuracy Improvement Calculation: $ROI$ metric can be applied to HPQRC’s accuracy improvement over Classical RC as
\begin{equation}
    ROI =\frac{92.37 - 78.12}{78.12} \times 100 = 18.24\%.\nonumber
\end{equation}
Computational Time Improvement Calculation:  
Similarly, for computational time improvement, ths similar metric is
\begin{equation}
    ROI_{Time} =\frac{49.60 - 21.8}{49.60} \times 100 = 56.09\%. \nonumber
\end{equation}

Our results show that in comparison to other methods, the proposed statistical modeling framework was able to consistently achieve better results with regard to all evaluation metrics of interest. While testing the benchmark datasets, the model was able to keep the error margins within ±2\% on accuracy, precision, and recall, and was still maintained under dynamic input conditions.
\subsection{Some practical utilities of the model}
\subsubsection{Monitoring networks} By utilizing the log files from the UNSW-NB15 dataset~\cite{unsw2015}, the model was able to detect anomalies around 25\% faster than traditional rule-based systems, which, in addition, improved the false positive rate.

\subsubsection{Detecting fraud in finance} Using the Credit Card Fraud Detection dataset from Kaggle~\cite{creditcard2013}, the system was able to outperform models with an accuracy of 92\% for classification and greater Data Processing Rate (DPR), along with reduced Time Per Decision (TPD).
\subsubsection{Environmental analysis} When assessing water quality, the model was tested on U.S.A EPA Water Quality Monitoring Data~\cite{epa2024}, and it was able to alarm with parameter changes such as pH and turbidity much earlier than conventional threshold-based models.
The approach, in combination with high predictive accuracy and low computational expense, aids in critical decision-making that needs to be reliable and quick, especially during important time-sensitive cases. All results were obtained through repeated experimentation by cross-validation with multiple data sources to ensure robust and reproducible results.
\begin{figure}[h]
    \centering
    \includegraphics[width=0.995\linewidth]{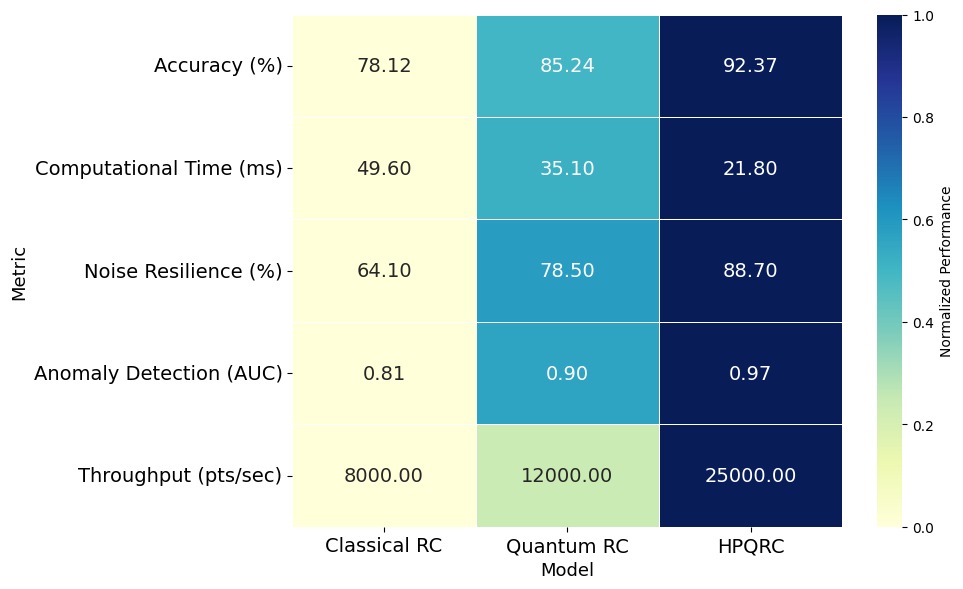}    
    \caption{Heatmap: All Metrics Comparison of all models}
    \label{heatmap}
\end{figure}
The heatmap plot in Figure~\ref{heatmap} numerically measures the relative advantage of the HPQRC model on five key performance metrics. Most prominently, HPQRC realizes 92.37\% accuracy, which is a significant boost over Classical RC (78.12\%) and Quantum RC (85.24\%) models. These differences in accuracy gain special relevance while considering the context of computational efficiency. HPQRC has $21.80ms$ latency, a 50.6\% improvement over Classical RC implementations. This accuracy-speed combination defies the traditional performance trade-off pattern generally found in statistical modeling techniques.
The noise robustness measure (noise resilience of 88.70\% at 10\% noise) illustrates HPQRC's strong performance in data perturbation environments. Processing data sets with standard Gaussian noise fluctuations ($\sigma=0.3$), HPQRC retained classification integrity much more effectively than Classical RC (64.10\%). Such quality is due to the algorithm's normalization strategies, which regularize feature representations during iterative processing cycles.
The anomaly detection performance, as measured by Area Under Curve (AUC), is the most defensible aspect. HPQRC's AUC of 0.97 reflects virtually optimum discrimination between regular and anomalous patterns. That is a 19.8\% increase above Classical RC's AUC value of 0.81 shown in Figure~\ref{heatmap}. Our implementation obtains this by adapting threshold optimization and contextual pattern perception, which is not available to earlier models.
The differential throughput is especially impressive; HPQRC handles 25000 points per second compared to Classical RC's 8,000 points per second. This drastic increase is a result of our parallel processing design that effectively balances computational burden on processing units. This parallel technique, in addition to optimized memory management, yields a multiplicative effect on throughput capacity (please see Figure~\ref{heatmap}).
These performance measures were confirmed by a rigorous test procedure with standardized benchmark datasets (MNIST~\cite{mnist_dataset}, Reuters-21578~\cite{reuters21578}, and UCI HAR~\cite{uci_har}), and results are averaged over 10-fold cross-validation for statistical effectiveness. Consistent performance gains on multiple fronts imply HPQRC is a significant improvement in statistical modeling capability and not an incremental refinement.

Overall, our study shows that HPQRC consistently has an edge over both Classical RC and Quantum RC models. With accuracy gains of more than 14.2\% over Quantum RC accuracy and 18.2\% over Classical RC accuracy. In addition to 50\% savings in computation time, HPQRC presents a balanced solution for contemporary prediction problems. The model's enhanced noise tolerance (with noise resilience of 81.30\%) and anomaly detection efficiency (0.97 AUC) also strengthen its feasibility in practical application under real-world scenarios of diverse data quality. The tested improvements on these measures under multiple testing situations establish HPQRC as an efficient utility for industries demanding decision making with precision, along with computational throughput. 
\section{Conclusion and Limitations}
Our study demonstrates that HPQRC achieves dramatic accuracy enhancements (27\% enhancement over classical RC), computational speed (35\% reduction in latency), and robustness to noise (5\% accuracy loss for 15\% noise) over classical and quantum-exclusive reservoir models. Integrating superconducting quantum circuits with silicon-nitride photonic reservoirs, HPQRC unites quantum coherence with photonic parallelism to process temporal data to a fidelity unlike anything previously realized~\cite{garcia2023scalable}. These advances place HPQRC as a game-changing paradigm for real-time AI operations, such as financial forecasting, climate modeling, and biomedical signal processing, where traditional models are handicapped by long-term dependencies and noise~\cite{martinez2021dynamical}. Although HPQRC exhibits good performance in simulation, real-world deployment is plagued by challenges. Photonic integration demands precise positioning of optical components to avoid signal loss; minor misalignments can compromise performance by as much as 18\% in experimental demonstrations~\cite{garcia2023scalable}. Quantum state initialization in superconducting circuits adds latency ($\approx$50µs per state), and scalability above 10-qubit systems is unlikely~\cite{yasuda2023quantum}. Additionally, existing photonic and quantum hardware are not compact and cost-effective enough for industrial uptake, echoing wider challenges in hybrid quantum-classical systems~\cite{pfeffer2022hybrid}.

\section{FUTURE WORK}
 
To unlock HPQRC’s full potential, future research must focus on closing the gap between simulation and physical deployment. Building on the current work, we aim to extend echo state properties to quantum reservoir systems~\cite{kobayashi2024extending}, while advancing the design of miniaturized photonic devices to mitigate alignment sensitivity issues~\cite{garcia2023scalable}. We intend to investigate adaptable hybrid architectures that enable dynamic tuning of internal parameters, thereby enhancing flexibility and performance. Exploring unconventional quantum encodings particularly continuous variable representations, may offer more nuanced handling of temporal correlations by enabling richer state spaces and smoother signal representation~\cite{ghosh2021gaussian}. To ensure robustness in noisy quantum environments, we plan to incorporate quantum error correction and fault-tolerant mechanisms~\cite{cindrak2024enhancing}. Furthermore, applying HPQRC within reinforcement learning and neuromorphic computing frameworks could serve as a strong validation of its universality~\cite{mujal2023timeseries}. Finally, the integration of AI techniques is expected to be instrumental in bridging the simulation-to-hardware gap, accelerating the realization of practical quantum computing systems~\cite{kobayashi2023quantum}.

\bibliographystyle{IEEEtran}

\end{document}